\documentclass{svjour3}                     % onecolumn (standard format)
\smartqed  % flush right qed marks, e.g. at end of proof
\usepackage{graphicx,amsmath}
\journalname{General Relativity and Gravitation}
\begin{document}
\title{A solution for galactic disks with Yukawian gravitational potential}
\titlerunning{Galactic disks with Yukawian gravitational potential}
\author{J. C. N. de Araujo \and O. D. Miranda}
\authorrunning{de Araujo \& Miranda} % if too long for running head

\institute{J. C. N. de Araujo \and O. D. Miranda \at
    Divis\~ao de Astrof\'\i sica \\
    Instituto Nacional de Pesquisas Espaciais  \\
    Avenida dos Astronautas 1758 \\
    S\~ao Jos\'e dos Campos - 12227-010 SP - Brazil \\
              Tel.: +55-12-39457200\\
              Fax: +55-12-39456811\\
              \email{jcarlos@das.inpe.br, oswaldo@das.inpe.br}
}

\date{Received: date / Accepted: date}

\maketitle

\begin{abstract}
We present a new solution for the rotation curves of galactic disks
with gravitational potential of the Yukawa type. We follow the
technique employed by Toomre in 1963 in the study of galactic disks
in the Newtonian theory. This new solution allows an easy comparison
between the Newtonian solution and the Yukawian one. Therefore,
constraints on the parameters of theories of gravitation can be
imposed, which in the weak field limit reduce to Yukawian
potentials. We then apply our formulae to the study of rotation
curves for a zero-thickness exponential disk and compare it with the
Newtonian case studied by Freeman in 1970. As an application of the
mathematical tool developed here, we show that in any theory of
gravity with a massive graviton (this means a gravitational
potential of the Yukawa type), a strong limit can be imposed on the
mass ($m_{\rm g}$) of this particle. For example, in order to obtain
a galactic disk with a scale length of $b\sim 10\, {\rm kpc}$, we
should have a massive graviton of $m_{\rm g}<< 10^{-59}{\rm g}$.
This result is much more restrictive than those inferred from solar
system observations. 
% 04.50.+h Gravity in more than four dimensions, Kaluza-Klein theory,
% unified field theories; alternative theories of gravity (see also
% 11.25.Mj Compactification and four-dimensional models)
% 98.62.Hr Spiral arms and bars; galactic disks
\end{abstract}

\section{Introduction}
\label{intro}

General relativity agrees, as is well known, with Newtonian
gravitation in the weak field limit (correspondence principle). Some
alternatives to general relativity theory, however, not necessary
respect this principle. In this way, independent of the specific
details of the models or theories (for example, scalar-tensor
theories of gravity, nonsymmetric gravitational theory, etc.), most
of them converge to the same weak field limit: the gravitational
potential is Yukawa-like (e.g., Refs.
\cite{moffat96,piazza03,rodriguez-meza05,sign05}).

A first motivation of the present paper is to present a mathematical
tool that can be used, independently of the particular theory of
gravity, for the study of the rotation curves of galaxies. In this
case, there is only one constrain: the potential must be
Yukawa-like.

Now, let us start recalling that the Newtonian potential $\phi$, as
is well known, follows the Poisson equation, namely
\begin{equation}\label{pe}
\nabla^{2}\phi=-4\pi G\rho.
\end{equation}
As a result, the potential of a point mass $m$ at a distance $r$
reads
\begin{equation}\label{np}
\phi=-\frac{Gm}{r},
\end{equation}
\noindent where $G$ is the gravitational constant.

Considering a gravitational potential of the Yukawa-type (hereafter
named Yukawian gravitation potential), we have for a point mass $m$
a potential given by
\begin{equation}\label{pm}
    \phi=-\frac{Gm}{r}e^{-r/\lambda},
\end{equation}
\noindent where $\lambda$ is a constant. Note that, for $\lambda
\rightarrow \infty$ this potential becomes identical to the
Newtonian one.

The field equation in this case reads
\begin{equation}\label{fe}
\left(\nabla^{2}-\frac{1}{\lambda^{2}}\right)\phi=-4\pi G\rho.
\end{equation}

Recall that the constant  $\lambda$ that appears in the Yukawa
potential is the Compton wavelength of the exchange particle of
mass $m_{\rm g}$. Using this interpretation for the present case
one can think of the exchange particle is a massive graviton. In
particular, the second motivation of the present paper is to
verify if it is possible to constrain the mass ($m_{\rm g}$) of
this particle.

In section 2 we present the new solution for the Yukawian
gravitational potential for a thin disk and compare it, in section
3, with the Newtonian solution obtained by \cite{free70}. In section
4 we present the main results and conclusions.

\section{The rotation curve for a thin exponential disk for a
Yukawian gravitational potential}

Toomre \cite{toom63} showed that the potential of a thin disk can be
written as follows
\begin{equation}\label{gpt}
\phi(r,z)=2\pi G \int^{\infty}_{0} J_{0}(k\,r)S(k)e^{-k|z|}dk,
\end{equation}
\noindent where $S(k)$ is related to the surface density, $\mu(r)$,
through the following  Bessel integral
\begin{equation}\label{}
\mu(r)=\int^{\infty}_{0}  J_{0}(k\,r)S(k)kdk,
\end{equation}
\noindent where
\begin{equation}\label{}
S(k)=\int^{\infty}_{0}  J_{0}(k\,u)\mu(u)udu
\end{equation}
\noindent comes from the Fourier-Bessel integral theorem.

It is straightforward to show that Eq.~(\ref{gpt}) satisfies the
Newtonian field Equation.

Once $\mu(r)$ is given one can obtain the potential of the disk, and
from the centrifugal-equilibrium condition, namely,

\begin{equation}\label{}
g(r)=\frac{v^{2}(r)}{r}=-\left(\frac{\partial \phi}{\partial r}
\right)_{z=0},
\end{equation}

\noindent one obtains the rotation curve of the disk.

Freeman (\cite{free70}) applied the above equations to obtain the
rotation curve for an exponential disk, whose surface density,
$\mu(r)$, in cylindrical coordinates is given by
\begin{equation}\label{sd}
\mu(r)=\mu_{0}e^{-r/b},
\end{equation}

\noindent where $b$ is the scale length of the disk.

It is worth stressing that the motivation for adopting an
exponential disk comes from observations. Surface photometry shows
that the two main components of most spiral and SO galaxies are
spheroidal and disk components.

In particular, the radial surface brightness distribution of the
disk follows an exponential law, namely, $I(r) = I_{0}e^{-r/b}$.
This implies that the surface density is exponential too, as we are
considering.

Finally, the circular velocity, i.e., the rotation curve, for this
exponential disk in the Newtonian gravitation reads
\begin{equation}\label{nrc}
v^{2}(r)= \pi G \mu_{0} b \left(\frac{r}{b}\right)^{2} (I_{0}K_{0}-
I_{1}K_{1}),
\end{equation}
\noindent where I and K are the modified Bessel functions, which are
calculated at ${r}/{2b}$.

We now consider how to get a solution for the rotation curve of a
thin disk for a Yukawian potential, using the Toomre approach.

The Yukawian potential for a thin disk can be obtained by rewriting
Eq.~(\ref{gpt}) properly, namely,
\begin{equation}\label{gptv}
\phi(r,z)=2\pi G \int^{\infty}_{0}
Z_{0}(\sqrt{|k^{2}-\lambda^{-2}|}\;r)S(k)e^{-k|z|}dk,
\end{equation}
\noindent where
%
%\begin{eqnarray}
\begin{equation}
  Z_{0}(\sqrt{|k^{2}-\lambda^{-2}|}\; r)=
\begin{cases}
  J_{0}(\sqrt{|k^{2}-\lambda^{-2}|}\; r) \qquad {\rm for} \qquad k \geq \lambda^{-1}
\cr I_{0}(\sqrt{|k^{2}-\lambda^{-2}|}\; r) \qquad {\rm for} \qquad k
< \lambda^{-1}
\end{cases}
\end{equation}
%\end{eqnarray}

As in the Newtonian case the Bessel functions appear in the solution
presented above. To account for the extra term ``$\phi /
\lambda^{2}$" in the Poisson equation the argument of the Bessel
functions must now contain ``$\lambda$".

Obviously, the direct substitution of the above equations satisfy
the Yukawian gravitational potential Eq.~(\ref{fe}).

We now apply the above equations to obtain the rotation curve for an
exponential disk, whose surface density, $\mu(r)$, in cylindrical
coordinates is given by Eq.~(\ref{sd}).

After some manipulation and an eventual change of variables the
rotation curve reads
\begin{eqnarray}\label{vrc}
v^{2}(r) &=& 2\pi G \mu_{0} r \nonumber
\\ & &\times \Biggl[\int_{b/\lambda}^{\infty}
\frac{\sqrt{x^{2}-b^{2}/\lambda^{2}}}{(1+x^{2})^{3/2}}\,
J_{1}\left(\frac{r}{b}\sqrt{x^{2}- b^{2}/\lambda^{2}} \right) dx
\nonumber \\ & & - \int^{b/\lambda}_{0}
\frac{\sqrt{b^{2}/\lambda^{2}-x^{2}}}{(1+x^{2})^{3/2}}\,
I_{1}\left(\frac{r}{b}\sqrt{b^{2}/\lambda^{2} - x^{2}}\right)
dx\Biggr]
\end{eqnarray}
It is worth stressing that we have parameterized the above equation
in terms of the ratio ``$b/\lambda$", which helps one to analyze how
different a theory with a Yukawian gravitational potential is as
compared to the Newtonian theory.

Note that for $b/\lambda \ll 1$ the second integral of the
Eq.~(\ref{vrc}) goes to zero and the first one becomes
\begin{equation}
\int_{0}^{\infty}
\frac{x\,J_{1}\left(\frac{r\,x}{b}\right)}{(1+x^{2})^{3/2}}\,
 dx = \frac{1}{2}\,\frac{r}{b} \,(I_{0}K_{0}-I_{1}K_{1})\, ,
\end{equation}
\noindent which substituting into Eq.~(\ref{vrc}) gives, as
expected, the Newtonian rotation curve.

In the next section we integrate numerically the Yukawian rotation
curve for different values of ``$b/\lambda$" and compare them with
the Newtonian rotation curve.

\section{Yukawian versus Newtonian disk}

We now compare the Newtonian rotation curve with those that come
from a Yukawian gravitational potential. To proceed, however, one
needs to integrate Eq.~(\ref{vrc}), which has three parameters to be
specified, namely, $\lambda$, $\mu_{0}$ and $b$.

The $\lambda$ parameter is in principle free, since we are not
considering any specific Yukawian gravitational theory. The
parameters $\mu_{0}$ and $b$ can be inferred from observations, as
already mentioned. For reasons that become clear later on we here
only need to know $b$.

There is in the literature a series of studies concerning the
determination of the disk parameters of spiral and SO galaxies. It
is worth noting that there is not any universal value to the scale
length $b$, different galaxies present different values for this
parameter. From these studies one can conclude that $b \sim 1 -10$
kpc \cite{flor93}.

\begin{figure}
\center\includegraphics[width=84mm]{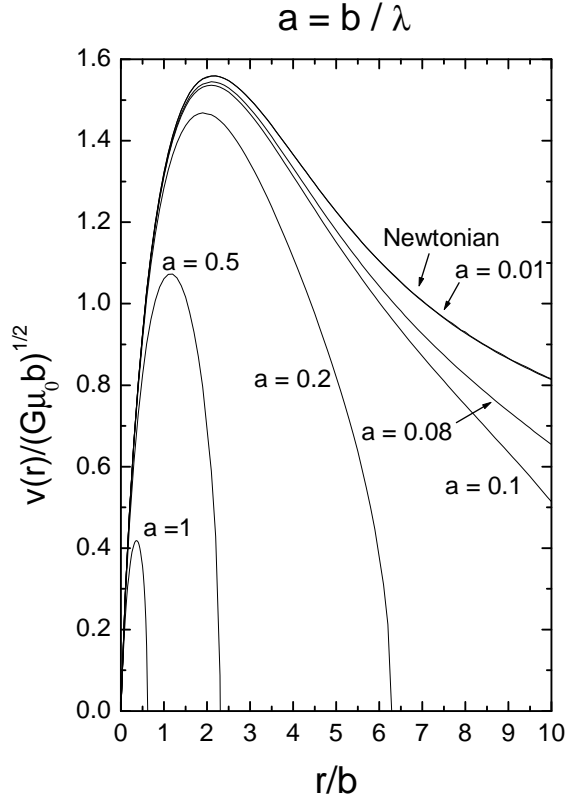}
\caption{\label{disc} Comparison between the Newtonian rotation
curve with different Yukawian rotation curves parameterized in
terms of ``$b/\lambda$".}
\end{figure}

Since $\lambda$ is a free parameter, whose value depends on the
particular theory adopted, we do not fix any particular value for
it. Instead, since the scale length is constrained by observations,
we present the rotation curves parameterized in terms of
``$b/\lambda$".

In Fig.~\ref{disc} we compare the Newtonian rotation curve with
different Yukawian rotation curves. We plot the velocity, in terms
of $\sqrt{G\, \mu_{0}\, b}$, which has dimension of velocity, versus
the distance to the center of the disk in units of the scale length,
$b$.

The Newtonian rotation curve has a maximum velocity around $r/b \sim
2$, and for larger values of $r$ the velocity decreases
monotonically. This is a well known result. On the other hand, the
behavior of the Yukawian rotation curves is the following: the
velocity increases up to a maximum value and then decreases; the
greater the ratio ``$b/\lambda$", the lower the maximum velocity is.
The size of the disk is strongly dependent on the value of the ratio
``$b/\lambda$". The greater the ratio ``$b/\lambda$", the smaller
the disk is. As expected, for $b/\lambda \ll 1$ the rotation curve
becomes Newtonian.

These behaviors of the Yukawian rotation curves strongly constrain
the value of $\lambda$. As a result, for theories where $\lambda$ is
the Compton wavelength, strong constrains can be imposed on the
putative mass of the graviton.

To the Yukawian rotation curve be approximately the Newtonian one,
we need $b/\lambda \ll 1$. Recall that the Compton wavelength,
$\lambda_{c}$, of a particle of mass $m_{\rm g}$ reads
\begin{equation} \nonumber
\lambda_{c}= \frac{h}{m_{\rm g}c},
\end{equation}

\noindent where $c$ is the velocity of light and $h$ is the Planck
constant. For $b = 10$ kpc we should have
\begin{equation} \nonumber
m_{\rm g} \ll 10^{-59} {\rm g}.
\end{equation}
%

%\section{Conclusions}
\section{Discussions and conclusions}

We present a new solution for the rotation curves of a thin disk for
a Yukawian gravitational potential. This thin disk can represent the
disks present in spiral and SO galaxies. It is important to stress
that the mathematical tool developed here can be used to study
rotation curves in any model or theories of gravity which produce a
gravitational potential of the Yukawa type.

In particular, we compare the Yukawian rotation curve with the
Newtonian one. The Newtonian rotation curve has a maximum velocity
and for larger values of $r$ the velocity decreases monotonically.

The main characteristics of the Yukawian rotation curves are the
following: the velocity increases to a maximum value and then
decreases; the greater the ratio ``$b/\lambda$", the lower the
maximum velocity is. The greater the ratio ``$b/\lambda$", the
smaller the disk is. As expected for $b/\lambda \ll 1$ the rotation
curve becomes Newtonian.

These behaviors of the Yukawian rotation curves strongly constrain
the value of $\lambda$. As a result, for theories where $\lambda$ is
the Compton wavelength of a exchange particle, say a graviton,
strong constraints can be imposed on its putative mass. For $b = 10$
kpc, for example, we find $m_{\rm g} \ll 10^{-59} {\rm g}$.

It is worth stressing that the best bound on the graviton mass from
planetary motion surveys is obtained by using Kepler's third law to
compare the orbits of Earth and Mars, yielding $m_{\rm g} < 10^{-54}
{\rm g}$ \cite{larsonhiscock}. With such a mass and for $b = 10$ kpc
one would obtain that $b/\lambda \sim 10^{5}$, which would imply
that galactic discs could not exit (see Fig.~\ref{disc}) for a
Yukawian gravitational potential with this value of $m_{\rm g}$.

A study of interest, concerning a Yukawian gravitational potential,
would be focused on a complete model for the rotation curve
including, besides a disk, a halo. Such a study might impose
additional constraints on $\lambda$ or on the putative mass of a
graviton.

Also interesting is to study clusters of galaxies in Yukawian
gravitational theory. Since the size of clusters extends to $\sim
1$ Mpc, strong limits may be imposed on $\lambda$ or on the mass
of the graviton. We leave these studies, however, to another
papers to appear elsewhere.

Although we leave these studies to another papers, our results
summarized by the $a$ parameter in Fig. 1, can already be compared
to the galaxy rotation curves of the sample studied by, for example,
Ref. \cite{brownstein}.

In general, the behavior of the rotation curves, which come from the
observations, is the following: the velocity increases as a function
of the distance $r$ to the center of the galaxy; it then becomes
nearly flat for large values of $r$.

The innermost part of the rotation curves, where the velocity is an
increasing function of $r$, is dominated by the galactic disk. This
very behavior allows us to compare our results with observations. On
the other hand, the outermost part of the rotation curve, where it
becomes nearly flat, is dominate by the galactic halo (see, e.g.,
\cite{peacock99}, page 371).

In particular, Brownstein and Moffat \cite{brownstein} analyzed the
rotation curves of a sample of 101 galaxies compiled from
\cite{begeman,sanders,deblok,verheijen,sofue,romanowsky}. All these
galaxies present a behavior, in the innermost part of their rotation
curves, which are consistent with a Newtonian gravitational
potential. There is, as a result, a perfect agreement between our
results and the observations if $a \ll 1$. This very fact, as
already mentioned, strongly constrain the value of the graviton
mass.

Moreover, numerous studies of the Tully-Fisher (TF) relations and
their applications have been conducted in the past. In most cases
their aim was to find observables that reduce the scattering and
thus improve this relation as a tool for measuring distances to
spiral galaxies, and also to gain insights in the formation and
structure of galaxies \cite{hinz}.

In particular, \cite{verheijen} investigate the statistical
properties of the Tully-Fisher (TF) relations for a volume-limited
complete sample of spiral galaxies in the nearby Ursa Major Cluster.
They concluded that the TF relation reflects a fundamental
correlation between the mass of the halo and the total baryonic mass
of the galaxies. This kind of result certainly reinforces our
interest to apply the formalism here discussed on a more complete
model for rotation curves which includes a massive dark halo.

\begin{acknowledgements}
JCNA would like to thank the Brazilian agency CNPq for partial
support.
\end{acknowledgements}

\end{document}